\documentclass[aps,prx,reprint]{revtex4-2}

\usepackage{graphicx}
\graphicspath{C:\Users\Albert\Documents\Research - NIST\Publications and Presentations\WSF-1 Weyl Fermions\arxiv\Drafts}
\usepackage{epstopdf}
\usepackage{dcolumn}
\usepackage{bm}
\usepackage{amsmath}
\usepackage{mathrsfs}
\usepackage{amssymb}
\usepackage{wasysym}
\usepackage{color}
\usepackage{amssymb}
\usepackage{dsfont}
\usepackage{comment}
\usepackage{bigints}
\usepackage{setspace}

\begin{document}

\title{A Non-Topological Approach to Understanding Weyl Semimetals}

\author{Antonio Levy$^{1}$}
\email{antonio.levy@nist.gov}
\author{Albert F. Rigosi$^{1}$}
\author{Francois Joint$^{2}$}
\author{Gregory S. Jenkins$^{3}$}
\affiliation{$^1$National Institute of Standards and Technology, Gaithersburg, Maryland 20899, USA}
\affiliation{$^2$Department of Physics, University of Maryland, College Park, MD 20742, USA}
\affiliation{$^3$Laboratory for Physical Sciences, College Park, MD 20740, USA}

\begin{abstract}
In this work, chiral anomalies and Drude enhancement in Weyl semimetals are separately discussed from a semi-classical and quantum perspective, clarifying the physics behind Weyl semimetals while avoiding explicit use of topological concepts. The intent is to provide a bridge to these modern ideas for educators, students, and scientists not in the field using the familiar language of traditional solid-state physics at the graduate or advanced undergraduate physics level.
\end{abstract}

\date{\today}

\maketitle


\section{Introduction}

Weyl fermions (defined below) have historically been of interest in answering fundamental questions about the universe, particularly the observation of the matter-antimatter imbalance.\textsuperscript{1} The family of elementary particles classified as fermions, or particles of half-integer spin, are important in the Standard Model that unifies three of the four known forces of nature. Within the model are twenty-four families of fermions. Almost all of them are massive \textit{Dirac fermions}. Within the family of \textit{Dirac fermions} lies a subset class known as \textit{Weyl fermions}, the set of fermions that are massless. Those well-versed in the physics of the weak nuclear force will recall that those interactions are stronger for both left-chiral matter and right-chiral antimatter than their counterparts of opposite chirality (chirality summary shown in Fig. 1). 

Weyl and 3D Dirac semimetals are the only systems in which signatures of Weyl fermions have been observed. Although they are not fundamental particles, the excitations in these material systems offer a unique playground to study Weyl fermion physics like the chiral anomaly.\textsuperscript{2-6} Novel properties predicted for Weyl semimetals (WSMs) like significantly reduced scattering, Drude enhancement along applied magnetic fields, and long-lived spin-polarized currents in the presence of magnetic fields could lead to myriad applications in fields like spintronics and quantum computing. 

Most introductions to WSMs and Weyl Fermions are mathematically intense, making it difficult to develop an intuitive understanding of their physics. There are a few works that try to explain the concepts behind Weyl fermions in an educational context.\textsuperscript{7-9} This paper aims to provide a conceptual introduction accessible to educators, students, and scientists who have some understanding of traditional condensed matter physics. 

\section{Background on Weyl Fermions}

The concept of Weyl fermions originated from the fusion of two familiar topics. The first is the energy-momentum relation from relativity expressed as: $E^{2}= \left( pc \right) ^{2}+ \left( mc^{2} \right) ^{2}$. The second is the time-dependent Schrödinger equation from quantum mechanics expressed as: $-\frac{\hslash^{2}\triangledown ^{2} \psi }{ \left( 2m \right) }=i\hslash\frac{ \partial  \psi }{ \partial t}$.

To incorporate relativity into a quantum mechanical formula, Dirac treated each quantity in the energy-momentum relation as an operator acting on a wavefunction. Using $p=-i\hslash\triangledown$ and $E=i\hslash\frac{ \partial }{ \partial t}$, the resulting equation is: 

\begin{equation}
  \left( -\frac{1}{c^{2}}\frac{ \partial ^{2}}{ \partial t^{2}}+\triangledown ^{2} \right)  \psi =\frac{m^{2}c^{2}}{\hslash^{2}} \psi
\end{equation}

This is known as the Klein-Gordon equation. It does not include spin and is therefore applicable to zero-spin bosons. 

Dirac transformed this equation by taking the square root of the operators, which requires consideration of all three spatial dimensions and the time dependence. The transformation involves 4 $\times$ 4 matrices, called the Dirac Gamma matrices, that are directly related to the Pauli matrices $\overrightarrow{ \sigma }$ that describe half-spin fermions. The Dirac equation can be written in matrix block form with a particle-hole (\textit{p-h}) – or particle-antiparticle – basis:

\begin{equation}
\left[\begin{array}{cc}
  \left( \frac{\varepsilon}{c}-mc\right) & -\overrightarrow{p} \cdot \overrightarrow{\sigma} \\
  -\overrightarrow{p} \cdot \overrightarrow{\sigma} &  \left( -\frac{\varepsilon}{c}+mc\right)
  \end{array}\right]  
\left[ \begin{array}{cc}
 \psi _{p}\\
 \psi _{h}
\end{array}
 \right]=0,
\end{equation} 

where $\varepsilon$ is the energy of the particle. On the other hand, rather than using a particle-hole basis, the Weyl equation uses left- and right-chiral particles as the basis:

\begin{equation}
\left[\begin{array}{cc}
 mc & \frac{\varepsilon}{c}-\overrightarrow{p} \cdot \overrightarrow{\sigma} \\
  \frac{\varepsilon}{c}+\overrightarrow{p} \cdot \overrightarrow{\sigma} &  mc
  \end{array}\right]  
\left[ \begin{array}{cc}
 \psi _{L}\\
 \psi _{R}
\end{array}
 \right]=0,
\end{equation} 

\begin{figure}[h]
  \centering
  \includegraphics[width=0.48\textwidth]{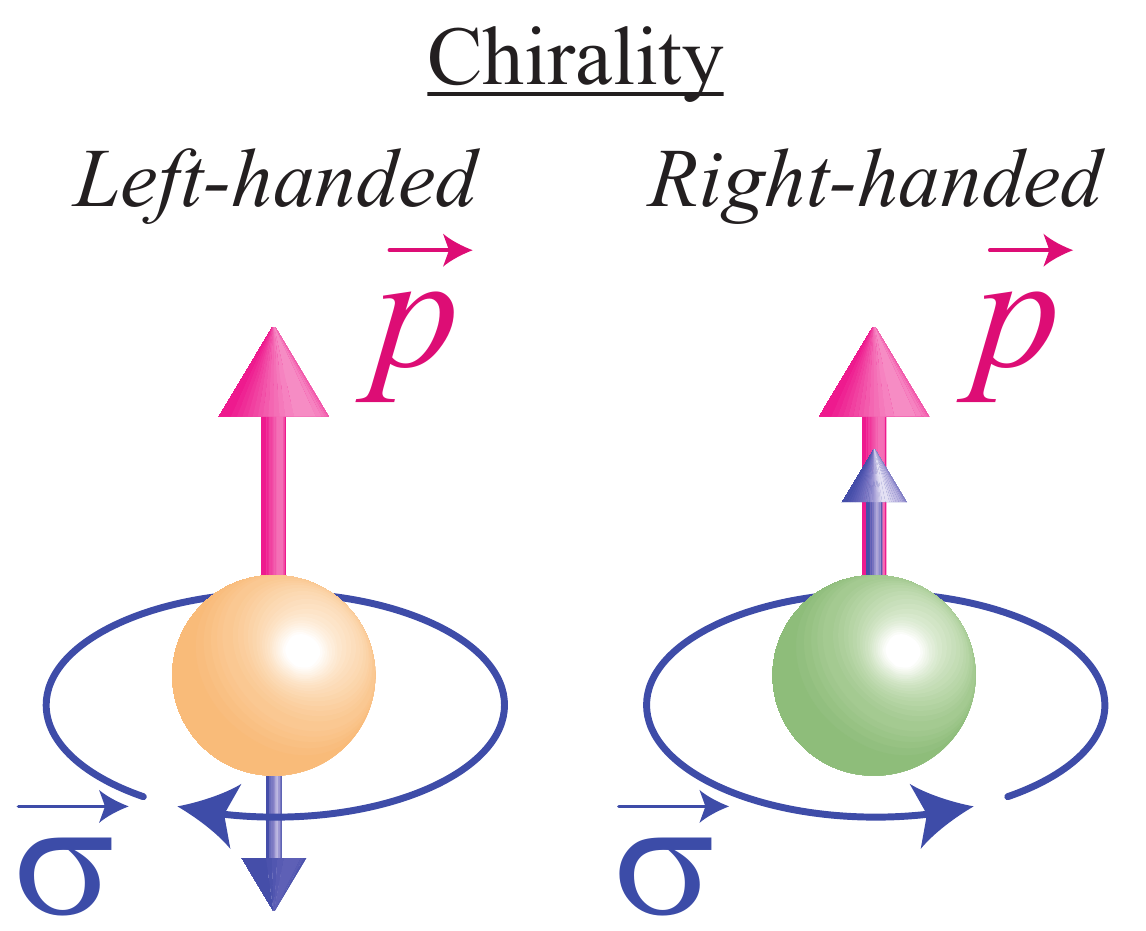}
  \caption{A left-handed chiral particle has a spin $\sigma$ that is antiparallel to its momentum $p$. The spin and momentum are parallel for a right-handed chiral particle.}
 
\end{figure}

Massive particles in the Weyl equation prevent pure chiral eigenstates from emerging out of Equation (3). Pure chiral particles must be massless. The eigenstates of massless Weyl particles only involve the momentum and a parallel or antiparallel spin as shown in Figure 1.

In solid state physics, the dispersion relation is a mathematical description of the available energies for an electron moving about in a material with a given momentum. It is frequently plotted as energy versus momentum (also referred to as \textit{k}-space). Further details on the basics of solid state physics can be found in standard textbooks.\textsuperscript{10} The relativistic energy-momentum dispersion relation for massless particles is  $E=pc$ . For particle-like excitations (or quasiparticles) moving through solid materials, the velocity can be a constant other than the speed of light while still obeying the Weyl equation. Casting the momentum in terms of wavenumber  \( \overrightarrow{p}=\hslash \overrightarrow{k} \) , the operative Weyl dispersion relation becomes $E=\hslash kv$, where the  $v$  is the Fermi velocity which plays a role similar to  $c$ in the relativistic equations. The zero-energy is defined at zero momentum and is called the \textit{Weyl point}, and its presence in Weyl semimetals is responsible for the novel physics they are predicted to exhibit.

\section{Chiral Anomalies}

To appreciate how a chiral anomaly emerges, it will help to recall a few concepts from statistical and solid-state physics. Within any material, electrons and other quasiparticles only have access to a limited number of momenta. Mapping out these allowable momenta creates what is commonly referred to as \textit{k}-space (closely related to the reciprocal lattice of a crystalline solid). The highest-energy electrons, in some sense, define the bounds of achievable momenta, and these bounds sketch out an object in \textit{k}-space called a Fermi surface. The Fermi surface for a Weyl particle is a sphere centered on the zero-momentum Weyl point.

Due to quantum mechanics, the allowable energies at which electrons are permitted to exist are discrete (and by extension, are also discrete in \textit{k}-space via Fourier transformation). These quantized energies make up the band structure of a crystal. For the remainder of this paper, it is assumed that the Fermi energy exists solely within the Weyl band and that other bands are sufficiently separated in energy so as not to perturb the Weyl state. With this last concept recalled, we are ready to describe the anomaly.

Consider the divergence theorem as applied to a charge \textit{Q}, charge density  $\rho$ and electrical current density \textit{j}: 

\begin{equation}
\frac{dQ}{dt}\mathrm{=} \iiint _{}^{} \left( \frac{ \partial  \rho }{ \partial t}+\overrightarrow{\triangledown }\cdot \overrightarrow{j} \right)  d^{3}x
\end{equation}

Since charge is conserved, the integral extends over the entire volume of the system, leaving both sides of Eq. (4) equal to zero. This expression indicates that the total amounts\ of charge entering and leaving the system are equal.  Combining Eq. (4) with Eq. (3) for \textit{m} = 0 under non-zero electromagnetic fields (which is done by replacing  \( \overrightarrow{p} \)  with  \( \overrightarrow{p}-\frac{e\overrightarrow{A}}{c} \) ), the difference between the number of right- and left-chiral particles,  \( n_{R}-n_{L} \)  can be obtained after considerable effort: 

\begin{equation}
\frac{d}{dt} \left( n_{R}-n_{L} \right) \mathrm{\propto} \iiint _{}^{}\overrightarrow{E}\cdot \overrightarrow{B} d^{3}x
\end{equation}

Equation (5) shows that if a population of Weyl fermions is subjected to non-zero applied electric and magnetic fields, the chirality of the population will change with time. This is equivalent to saying that Weyl fermions of a single chirality will be annihilated and replaced with Weyl fermions of the opposite chirality. This is the chiral anomaly.

The chiral anomaly in a crystal is best understood intuitively by considering the effects of applied magnetic and electric fields on the Fermi surface with linear dispersions illustrated in Fig. 2. The momentum direction of any quasiparticle on a spherical Fermi surface is radially outward. In the depicted generic WSM, there are two locations in \textit{k}-space around which the band structure obeys the Weyl equations. Each location hosts one species of chiral particle where the spin is either radially inward opposite the momentum (orange Fermi surface pocket on the left side of each subfigure) or purely aligned radially outward with the momentum (green Fermi surface pocket on the right side of each subfigure). As will be discussed later, including only a single chiral Fermi pocket violates the conservation of energy. 

So what happens to the Fermi surface if you apply a magnetic field? It will distort, expanding or contracting by an amount proportional to the dot product of the spin and the magnetic field (Zeeman effect). To intuitively understand these distortions, consider the left-handed chiral (orange) Fermi surface pocket in Figure 2(e-g) at the point where the spin of a quasiparticle is purely antiparallel with applied magnetic field. The quasiparticle’s energy will decrease with applied field. This decrease in energy, through the linear energy dispersion, causes the momentum to decrease. The decrease in momentum shifts this point of the Fermi surface toward the center of the Fermi pocket sphere. Likewise, the opposite point, where the field and spin are parallel, shifts to higher momentum away from the center. All the points where the spin is perpendicular to the field do not change their energy or momentum. Visualizing the smooth interpolation between these points results in a Fermi surface distorted into an egg-like shape with the major axis aligned with field as shown in Figure 2(e). 

Similar effects on the Fermi sea occur from the charge of the quasiparticle subjected to an applied electric field. In this case, all negatively charged quasiparticles, regardless of their spin or location on the Fermi surface, decrease their (vector) momentum along the direction of the electric field causing a net momentum shift of the Fermi surface.

With either an applied electric or magnetic field, the changes in a singly-chiral Fermi pocket induces a net momentum and therefore an electric current. This current also transports a net spin current. However, when both (oppositely) chiral spherical Fermi surface pockets are considered, the application of a magnetic field (with no electric field) causes two equal and oppositely-oriented distortions that negate the total current. In fact, this can be understood as the reason that all Weyl semimetals must include pairs of oppositely-chiral Fermi pockets; a static magnetic field should not be able to generate a perpetual current in a material system with a finite scattering rate like Weyl semimetals, as this would violate the conservation of energy. Furthermore, in order not to violate conservation of energy, this current would have to be superconducting. At high fields, there would be a strong current that would flow consistently in one direction, which would persist independent of the direction in which an external electric field was applied to the material. This would mean that carriers could flow in the direction opposite to the bias applied to the material, leading to negative power dissipation in the material as a result of an applied external electric field (provided that field was not large enough to fundamentally alter the band structure). 

To formulate an expression for the generated currents, the quasiparticles removed and added at various momenta with applied fields must be properly counted. Recall the density of states, which is the derivative of the number of states with respect to energy in the system. The density of electronic states (g, which we define as the number of states per unit energy at a specific point in \textit{k}-space) is energy dependent and therefore becomes dependent on k-space location ($\theta $, defined as the polar angle between the \textit{k}-vector and the applied fields) with the application of fields. 

Consider for simplicity the case with both fields pointing in the  $\mathrm{-}x$ direction (as in Fig. 2). Then, for the right-chiral location, we track the difference between right- and left-chiral particles $n_{R} (\theta)-n_{L} (\theta)$, which is proportional to integrated subtraction of the density of states, expressed as  $g_{R ( L ) } (  \varepsilon _{F} ) =g (  \varepsilon _{F}+ (-)a_{1}Bcos (  \theta  ) +a_{2}Ecos (  \theta  )  )  $ for the right and left pockets, where  $a_{1} $ and  $a_{2} $ are constants. The difference in sign in front of the  $a_{1}Bcos (  \theta  )  $ is due to the opposite spin of the two pockets, highlighting the antiparallel nature of the chirality. Approximating for small fields,  $g_{R ( L ) } (  \varepsilon _{F} )  $ can be rewritten:  $g_{R ( L ) } (  \varepsilon _{F} ) =g (  \varepsilon _{F} ) + ( - ) a_{1}Bcos (  \theta  ) g' (  \varepsilon _{F}+a_{2}Ecos (  \theta  )  )  $ or, equivalently, we could say: $g_{R} (  \varepsilon _{F} ) -g_{L} (  \varepsilon _{F} ) =2a_{1}Bcos (  \theta  ) g' (  \varepsilon _{F}+a_{2}Ecos (  \theta  )$.  The difference in the population of right- and left- chiral Weyl fermions is given by: 

\begin{equation}
\begin{aligned} \int _{0}^{ \pi } \left( g_{R} \left(  \varepsilon _{F}, \theta  \right) -g_{L}~ \left(  \varepsilon _{F}, \theta  \right)  \right) d \theta  \approx \\
2a_{1}B \int _{0}^{ \pi }\cos  \left(  \theta  \right) g^{'} \left(  \varepsilon _{F}+a_{2}Ecos \left(  \theta  \right)  \right) d \theta
\end{aligned}
\end{equation}

This subtraction is reduced to a derivative of the density of states for small fields. With additional symmetry considerations, the bounds of the integral are halved as long as we take into account  \( E \rightarrow -E \)  for the respective halves of the Fermi pockets:

\begin{equation}
\begin{aligned}
2a_{1}B \int _{0}^{ \pi }\cos  \left(  \theta  \right) g^{'} \left(  \varepsilon _{F}+a_{2}Ecos \left(  \theta  \right)  \right) d \theta  \approx \\
2a_{1}B \int _{0}^{\frac{ \pi }{2}} \left( g^{'} \left(  \varepsilon _{F}+a_{2}E\cos  \left(  \theta  \right)  \right) -g^{'} \left(  \varepsilon _{F}-a_{2}Ecos \left(  \theta  \right)  \right)  \right) \\
\cos  \left(  \theta  \right) d \theta \\
\propto 2a_{1}a_{2}E B \int _{0}^{\frac{ \pi }{2}}g^{''} \left(  \varepsilon _{F} \right) \cos ^{2} \left(  \theta  \right) d \theta~
\end{aligned}
\end{equation}

where the approximation  \( g' \left(  \varepsilon _{F} \pm a_{2}E\cos  \left(  \theta  \right)  \right) = g' \left(  \varepsilon _{F} \right)  \pm a_{2}E\cos  \left(  \theta  \right) g'' \left(  \varepsilon _{F} \right)  \)  is used to go from the second line of Eq. (7) to the third. 

With these considerations, the origin of Eq. (5) becomes apparent. Since the first time derivative of  \(  \left( n_{R}-n_{L} \right)  \)  is a nonzero quantity under the experimental conditions, a net chiral current, the $``$chiral anomaly,$"$  emerges and begins pumping particles from the left-chiral Fermi pocket to the right-chiral Fermi pocket leading to a net accumulation of right-chiral particles. Furthermore, the total number of right-chiral particles increases faster than the corresponding decrease seen in the left-chiral particles. For static fields, this ever-growing chirality imbalance is eventually counteracted by quasiparticle scattering between and within the two chiral Fermi pockets. A chirality-imbalanced equilibrium value is established that depends on both pumping rate and (current relaxation) scattering rate. 

\begin{figure}[h]
  \centering
  \includegraphics[width=0.48\textwidth]{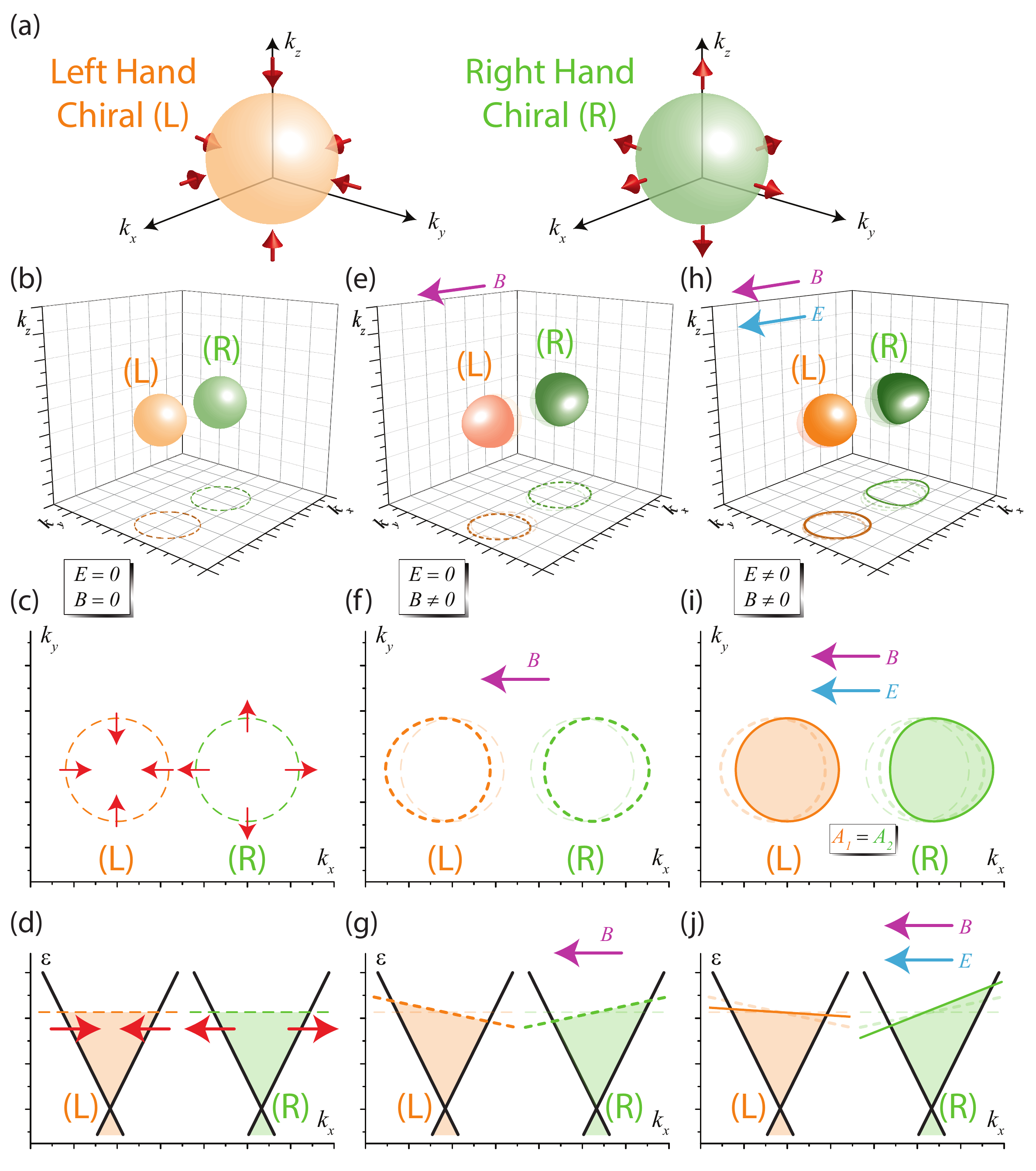}
  \caption{The chiral anomaly in Weyl semimetals is understood by distortions and shifts of the Fermi surface caused by the application of electric and magnetic fields. (a,b) With no applied field, the Fermi surface is two spheres: one Weyl pocket is left-handed chiral (orange) with spins perpendicular to the Fermi surface (pointing inward) and the other right-handed chiral (green) with spins (red arrows) perpendicular to the Fermi surface (pointing outward). (c) The  $ k_{x}\mathrm{-}k_{y}~ $ planar cut through the center of the Fermi pocket spheres results in two circles that demarcate allowable quasiparticle momenta. (d) The energy of the quasiparticles, with momenta taken as a  $ k_{x}\mathrm{-} $ line cut through the centers of the circles, is given by the linear dispersion  $E=\hslash k_{x}v $  (where velocity  $ v $  is constant). (g) Application of an external magnetic field in the - $ x $  direction raises or lowers the energy of the quasiparticle depending on the direction of the spin. Changes in energy are connected to changes in momenta through the linear energy dispersion resulting in (e,f)  $ k $ -space shifts and distortions of the Fermi surface. (j) The addition of an applied electric field interacts with the (negative) charge of the quasiparticle causing a vectoral increase of the momenta in the - $ x $  direction that either increases or decreases the quasiparticle kinetic energy depending on the initial direction of the momentum. (h,i) The combined effects of the applied fields distort and shift the Fermi surface, generating currents that, in the depicted case, are largest for the (green) right-handed chiral carriers.}
 
\end{figure}

\section{Semi-Classical Picture of Drude Enhancement}
Another interesting phenomenon regarding WSMs is the existence of Drude enhancement. To get comfortable with this topic, one must recall the Drude model of AC conductivity from solid state physics: 

\begin{equation}
\sigma _{D} \left(  \omega  \right) =\frac{ \omega _{D}^{2}}{4 \pi  \left( \frac{1}{ \tau}-i \omega  \right) }
\end{equation}

As  \(  \omega  \rightarrow 0 \)  (when  \( B=0 \) ) the DC conductivity appears as  \(  \sigma _{DC}=\frac{ \omega _{D}^{2} \tau}{4 \pi } \) , where the term  \(  \omega _{D}^{2} \)  is referred to as the Drude weight. The Drude weight and conductivity of a material are properties that are generally well-described by the equations of motion for the charge carriers. 

For most metallic and semi-metallic systems in the semi-classical limit, application of a magnetic field parallel to the driving electric field does not affect current since there is no Lorentz force. In the case of Weyl and 3D Dirac semimetals, such an arrangement of applied fields is expected to increase the Drude weight and this was recently verified experimentally.\textsuperscript{11} Most explanations of this very unusual effect rely heavily on the mathematical consequences of the Berry’s phase curvature near the Weyl points. Such explanations, though rigorous in the semi-classical limit, do not provide an intuitive explanation of the origin of such an effect. 

Again, studying the Fermi surface of the Weyl state provides a non-topological approach to understanding this effect. Since the Weyl pockets are represented in \textit{k-}space as asymmetrically distorted (egg-shaped) Fermi surface pockets (Fig. 2 (h)), any applied magnetic field will have additional, distinct effects on both populations of chiral particles, pushing more of the particles into the portion of the Fermi surface that contributes to the net current, as shown in Figs. 2(h) and 2(i). The Drude weight is proportional to the current generated by the applied external electric field. Integrating over the Fermi surface of a pair of Weyl points in the semi-classical small-field limit, with  $\sqrt{\frac{2\hslash v_{x}v_{y}}{l_{B}^{2}}} \ll  \varepsilon _{F}$:

\begin{equation}
\begin{aligned}
\omega_{D}^{2} \propto \int d \Omega [ g_{+} (\varepsilon_{F},\overrightarrow{B}) +g_{-} (\varepsilon_{F},\overrightarrow{B})]  {v_{F}^{\parallel}}^{2} (\theta) \\
\propto \int _{}^{}d \Omega  \left[ g \left(  \varepsilon _{F} \right) +\frac{1}{2}g^{''} \left(  \varepsilon _{F} \right)  \left( a_{1}B\cos  \left(  \theta  \right)  \right) ^{2} \right] {v_{F}^{\parallel}}^{2} \left(  \theta  \right) 
\end{aligned}
\end{equation}

where  \( v_{F}^{\parallel} \left(  \theta  \right) =v_{F}\cos  \left(  \theta  \right)  \)  is the Fermi velocity component parallel to the applied fields and  $ l_{B}=\sqrt[]{\frac{\hslash c}{eB}}$  is the magnetic length. The second term in (9) yields the  \( B^{2} \)  Drude weight enhancement obtained using the Berry’s phase curvature.\textsuperscript{12}

\section{Quantum Mechanical Picture of the Chiral Anomaly and Drude Enhancement}

For a more comprehensive picture of Drude enhancement, we turn to the quantum limit of the Drude weight of a Weyl pocket in 3D. The energy dispersion of the \textit{n}\textsuperscript{th} excited Landau level (LL) for  \( B\parallel x \)  is given by:

\begin{equation}
\varepsilon _{n} \left( k_{x} \right) =\sqrt[]{\frac{ \left( 2n+ \left( 1+s \right)  \right) \hslash v_{y}v_{z}}{l_{B}^{2}}+\hslash^{2}v_{x}^{2}k_{x}^{2}}
\end{equation}

where  \( s= \pm 1 \)  correspond to the two spin eigenstates. Note that this approach can convert the 3D system into a series of 1D systems. Every LL (except  \( n=0 \) ) has an energy dispersion containing an effective mass. Therefore, only the  \( n=0 \)  Landau level is chiral, making it responsible for all of the nontrivial behavior associated with WSMs. Consider the chiral anomaly. For all  \( n \neq 0 \)  LLs, the application of an electric field simply shifts electrons within the LL. They do not change the population of the pocket or even the LL itself. 

\begin{figure}
  \centering
  \includegraphics[width=0.48\textwidth]{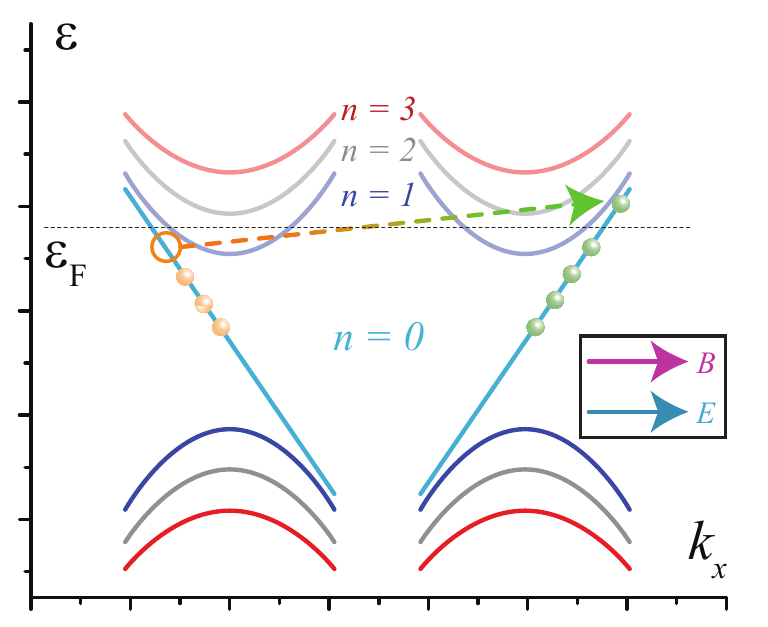}
  \caption{The energy dispersion of the various Landau levels along the magnetic field direction is illustrated. Left-handed particles are pumped from one Weyl pocket to the right-handed pocket. The formation of Landau levels makes the dispersion quasi one-dimensional, due to the discretization of kinetic energy perpendicular to the magnetic field (i.e., in the \textit{y} and \textit{z} directions).}
 
\end{figure}

Under an electric field parallel to the magnetic field (see Fig. 3), the number of carriers in one of the two  \( n=0 \)  LLs will decrease while the number in the other  \( n=0 \)  LL will increase. The dispersion in the two pockets’  \( n=0 \)  LLs are given by  \(  \varepsilon _{+} \left( k_{x} \right) =+\hslash v_{x}k_{x} \)  and  \(  \varepsilon _{-} \left( k_{x} \right) =-\hslash v_{x}k_{x} \)  for the left- and right-chiral cases, respectively. Applying an electric field along the \textit{x}-direction will deplete the population of one  \( n=0 \)  LL while increasing the population of the other  \( n=0 \)  LL by the same amount. 

The transfer of charge from one pocket to the other occurs entirely through the chiral  \( n=0 \)  LL of both pockets, since charge will be redistributed within each excited LL under an electric field. The amount of charge transfer is proportional to the electric field component parallel to the magnetic field and the degeneracy of the  \( n=0 \)  LL, which is proportional to  \( \frac{1}{l_{B}^{2}} \)  due to Landau quantization perpendicular to the magnetic field which, in our case, is the \textit{yz}-plane. From this, the chiral anomaly will persist at both low and high field strengths in WSMs. 

The spectral weight in the chiral  \( n=0 \)  LL is independent of the Fermi energy. From the quantum mechanical perspective, the increase in spectral weight with field can be understood as transferring more and more carriers from slower bands (where they are closer to the vertex) to the faster, massless  \( n=0 \)  LL. This quantum mechanical perspective on Drude enhancement gives a more nuanced understanding of the complexity of this effect without the need to rely on topological descriptions.

\section{Conclusions}

It is rather uncommon to find approaches to the physics behind Weyl fermions in the literature without involving a discussion on topology. Here, we explain the concepts of chiral anomalies and Drude enhancement in the context of Weyl semimetals intended for those not in the field. Non-topological approaches offer a more intuitive conceptual framework for understanding these physical systems. This understanding is useful when designing, performing, or analyzing data from experiments.

\end{document}